\documentclass[twocolumn,floatfix,pre]{revtex4}
      
 \usepackage{amsmath,amssymb}
 \usepackage{amsfonts}
 \usepackage{epsfig}

\begin{document}

\title{Emergence of Order in Selection-Mutation Dynamics}

\author{Christoph Marx}
\affiliation{Faculty of Physics, Universit\"at Wien,
 Boltzmanngasse 5, A-1090 Wien, Austria}
\author{Harald A. Posch}
\email{Harald.Posch@univie.ac.at}
\affiliation{Faculty of Physics, Universit\"at Wien,
 Boltzmanngasse 5, A-1090 Wien, Austria}
\author{Walter Thirring}
\email{Walter.Thirring@univie.ac.at}
\affiliation{Faculty of Physics, Universit\"at Wien,
Boltzmanngasse 5, A-1090 Wien, Austria}

\date{\today}
\begin{abstract}
We characterize the time evolution of a $d$-dimensional probability
distribution by the value of its final entropy. If it is near the maximally-possible
value  we call the evolution mixing, if it is near zero we say it is
purifying. The  evolution is determined by the simplest non-linear equation
and  contains a $d \times d$ matrix as input. Since we are not interested in a
particular evolution but in  the general features of evolutions of this
type,  we take the matrix elements as uniformly-distributed random numbers 
between zero and some specified upper bound.
Computer simulations show how the final entropies are
distributed over this field of random numbers. The result is that the
distribution crowds at the maximum entropy, if the upper bound is unity. 
If we restrict the dynamical
matrices to certain regions in matrix space, for instance to diagonal or
triangular matrices, then the entropy distribution is maximal  near zero,
and the dynamics typically becomes purifying.

\end{abstract}
\maketitle
\section{Motivation}
\label{sec_motive}
The abstract setting of evolution equations in fields like chemistry, biology, population dynamics, financial mathematics and similar subjects can be formulated as follows:
The state of the system is given by $d$ positive numbers $p_i$, which sum to unity and may be interpreted as probabilities, as relative populations, or financial assets. Their time dependence is governed by a dynamical equation, which expresses the time change of $p_i$ by the value of all the $p_j$. The basic question is whether in the course of time the state will purify  (which means that one 
$p_i$ will tend to unity and all the others to zero) or will become completely mixed (all $p_i$ become equal to $1/d$). To measure this tendency,  we use the quadratic entropy  \cite{Jumarie} $S$,
\begin{equation}
S =  \sum_{i=1}^d p_i (1 - p_i ) .
\label{entropy}
\end{equation}
$S$ vanishes for a pure state, it is $(d-1)/d$ for the completely mixed state, and is in between otherwise.
The evolution equations are constructed such that the basic properties of the $p_i$ are preserved and their values  asymptotically tend to a limit independent of the initial conditions. However, this limit will 
depend on the coefficients $\alpha_{ij}$ appearing  in the evolution equation as shown below. 
They are supposed to be determined by the accidental circumstances of the system and should be considered as random numbers. The basic issue therefore is, whether the asymptotic values for $S$, 
taken as a function of these coefficients, tend to cluster around zero,  or around $(d-1)/d$. Correspondingly,  we say that these evolution equations are either purifying or mixing. 

    We answer this question for the simplest type of evolution equation which, speaking financially, says the following: per unit time each player $i$ receives the fraction $\alpha_{ij}$ of the assets of the players $j$. Then, there is an overall tax such that the total amount of money remains unity. If we take all the 
$\alpha_{ij}$ to be random numbers between 0 and 1, we find that the  asymptotic values for $S$ 
cluster just below  $(d-1)/d$, and this distribution is the sharper the bigger the dimension $d$. 
Only for special classes of  $\alpha_{ij}$ they cluster near zero. This means that one can make a fortune only by an intelligent strategy, or that the ''survival of the fittest'' only holds under special circumstances. 

     The paper is organized as follows: In Section  \ref{sec_equation} we define our model and
outline the numerical procedure for the computation of the entropy $S$. In Section \ref{sec_results}
we summarize our numerical results for three to eight dimensional systems. The 
qualitative behavior is already apparent for the two-dimensional case, which may be solved
analytically. This is demonstrated in Section \ref{sec_theory}. We conclude in Section
\ref{sec_fitness}  with some remarks on fitness and the quantum-mechanical generalization.

\section{Dynamical evolution including mutation}
\label{sec_equation}

       We consider a $d$-dimensional dynamical system with a state vector 
\begin{equation}
       p=\{p_i\}, \: 0 \le p_i \le 1; \; i = 1,2,\dots,d , \nonumber       
\end{equation} 
which evolves according to 
\begin{equation}
\frac{\mathrm{d}p_i}{\mathrm{d}t}=\sum_{j=1}^{d}\alpha_{ij}p_j-p_i
         \sum_{j=1}^{d}\sum_{k=1}^{d}\alpha_{jk}p_k ~{.}
         \label{eq_fisher}
\end{equation}
The coefficients $\alpha_{ij}$ are elements of a $d \times d$ matrix $\alpha$ and are assumed 
to be strictly positive random variables with upper bounds specified below. 
From the general solution of Eq. (\ref{eq_fisher}),
\begin{equation}
      p(t)=\frac{\exp{(\alpha t)} p(0)}
      {\sum_{i=1}^{d}[\exp{(\alpha t)}  p(0)] _i} ~\mbox{,}
       \label{eq_sol}
\end{equation}
it follows that the vector $p$ is constrained to the simplex $S_d$,
\begin{equation}
    \sum_{i=1}^{d} p_i  = 1 \:,
\label{eq_simplex}
\end{equation}
provided the initial point is also contained in $S_d$, $\sum_{i=1}^d p_i(0) = 1$.
The state $p$ may be  regarded as  a $d$-dimensional probability distribution.  
The dynamics of Equation  (\ref{eq_fisher}) is closely related to the quasi-species equation 
introduced by Eigen and Schuster  \cite{EMS}, which has been extensively studied in the 
past \cite{Nowak}. It also corresponds to the evolution equation introduced by Nowak, 
Komarova and Niyogi for the study of language learning by children \cite{NKN}.
In terms of chemical reaction kinetics, it models evolutionary dynamics based on 
selection and mutation.

      $\alpha$ is considered a matrix and, therefore, is similar to an orthogonal sum of matrices 
in a Jordan normal form,
$$
\alpha = M^{-1} \sum_j (a_j + S_j) M,
$$
where the sum is over all Jordan blocks corresponding to different
eigenvalues  of $\alpha$ with degeneracy $d_j$, and  
$S$ is the shift,
\begin{equation}      
       S=\begin{pmatrix} 0 & 1 & 0 & 0 &  \cdots  \\ 0 & 0 & 1 & 0 &  \cdots  \\ 0 & 0 & 0 & 1 &   
       \cdots  \\
       \cdot & \cdot &  \cdot & \cdot & \cdots \\     
\end{pmatrix} .
\end{equation}
The matrix $a$ is diagonal in each $d_j \times d_j$ Jordan block. 
$\exp(a+S)$ can easily be calculated, because $a$ and $S$ commute and $S^d=0$. 
Since the off-diagonal elements of $\alpha$ are assumed here to be strictly positive, the Perron-Frobenius theorem applies. As a consequence, the maximum eigenvalue, $\lambda$, of $\alpha$ is 
always non degenerate and strictly positive, $\lambda > 0$. The  corresponding eigenvector, $v$,
is real and positive. The associated Jordan block is one-dimensional. It follows that the leading term
of  the numerator of Eq. (\ref{eq_sol}), for $t \to \infty$,  is $ e^{\lambda t} v $.  The exponential is cancelled by an analogous expression for the denominator.  We conclude from Eq. (\ref{eq_sol}) that  
the  asymptotic equilibrium  is determined by the eigenvector $v$ belonging to $\lambda$ and is normalized according to Eq. (\ref{eq_simplex}) and, thus, confined to the simplex $S_d$:
\begin{equation}
\lim_{t \to \infty} p(t) = v \left(\sum_{j=1}^d v_j\right)^{-1} .
\label{as_sol}
\end{equation}
  This asymptotic solution is simply denoted by $p$ for the remainder of this paper.            

        To quantify the asymptotic behavior of the system, we use the quadratic entropy defined in       
Eq. (\ref{entropy}),
which is bounded,
\begin{equation}
           0 \le S \le \frac{d-1}{d} \;.
\label{eq_S}
\end{equation}
The upper bound corresponds to a maximally {\em mixed} state, where all the $p_i$ are
equal, $p_i = 1/d$.  We say the dynamics is maximally mixing, if this state is
approached for $t \to \infty$.  The lower bound of $S$ corresponds to a {\em pure} state, where 
only one component of $p$ approaches unity and all the others vanish.
Accordingly, the dynamics is said to be maximally purifying, if such a state is asymptotically reached.
We say the system is (predominantly) mixing or purifying, if the distribution $\rho(S)$ is peaked at or near the respective upper or lower bound. 

\section{Results}
\label{sec_results}

In the following we consider classes of random matrices $\alpha$ with strictly positive elements, 
and determine the
asymptotic behavior of the system by computing the normalized entropy 
distribution $\rho(S)$, $\int_0^{(d-1)/d}\rho(S) dS = 1$, for each class.  This is achieved by numerically
diagonalyzing  (typically $10^7$ to $10^8$) reaction-rate matrices $\alpha$ and  constructing the 
entropy histogram from the re-normalized eigenvectors,  Eq. (\ref{as_sol}), belonging to the 
respective maximum eigenvalues.
For the classification of $\alpha$ we distinguish between i) diagonal elements (D),  ii) elements
above the  diagonal (U), and iii) elements below the diagonal (L).
The elements are randomly drawn from a uniform distribution such that an 
upper bound exists for each group:
\begin{eqnarray}
      \mbox{D:}& &0 \le  \alpha_{ii}  \le   D  \: ; \: 1\le i \le d  \nonumber \\
      \mbox{U:}& &0 < \alpha_{ij}  \le   U  \; ; \; 1 \le i < j \le d \\
      \mbox{L:}& &0 <\alpha_{ij}   \le   L  \; ; \; 1 \le j < i \le d  \nonumber\\
      \label{DUL}
\end{eqnarray}
We specify various classes of $\alpha$ by $D, U,$ and $L.$
\begin{figure}
\centering
   {\includegraphics[width=8cm]{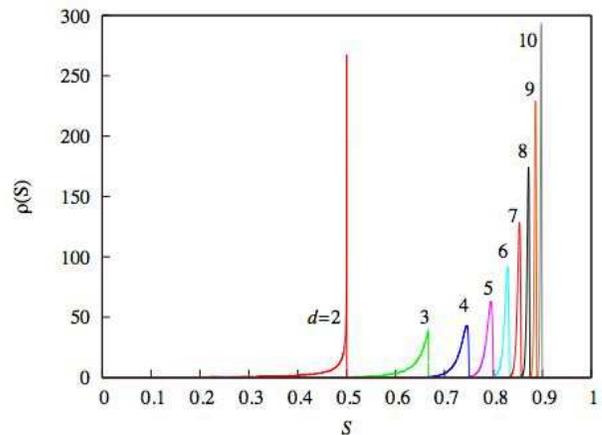}}
   \caption{(Color online) Random matrices with $D=U=L=1$: Normalized entropy
    distributions  $\rho(S)$  for various dimensions $d$ as specified by the labels.}
 \label{fig1}
\end{figure}

\subsection{ The most random case $D = U = L = 1$}

First we consider in Fig. \ref{fig1} the most random case, where the upper bounds for all
matrix elements are the same, $D=U=L=1$. The dimension $d$ is varied from 2 to 10.
Not surprisingly, the off-diagonal elements are responsible for effective mixing, and 
the distributions $\rho(S)$ are  sharply peaked
near their most-random edge, $(d-1)/d$. For $d \ge3$ they become narrower for increasing  $d$, and 
are expected to approach $\delta(1)$ for $d \to \infty$. 

\subsection{ The case $D=1$ and $U=L > 0$}
\begin{figure}
\centering
   {\includegraphics[width=8cm]{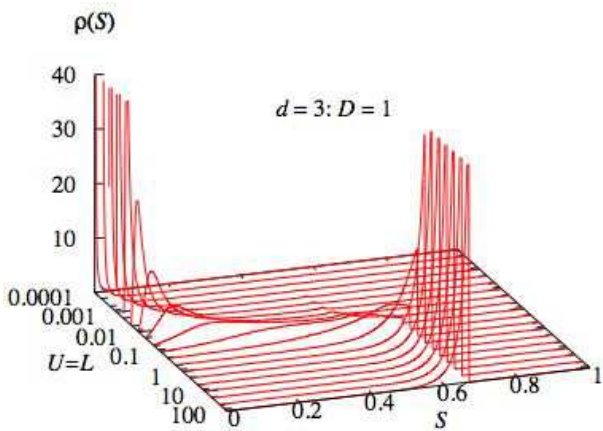}}
   \caption{(Color online)  Entropy distributions $\rho(S)$ in three dimensions $(d = 3)$ for 
   random matrices
   with $D=1$ and upper bounds for the off-diagonal elements $U = L$ varying between $10^{-4}$    
   and unity.  The distributions are clipped at 40 for display purposes. The selected off-diagonal
   bounds $U=L$ are arranged on a logarithmic scale.}
 \label{fig2}
\end{figure}

In Fig. \ref{fig2}  the diagonal elements for a three-dimensional system are bounded by 
unity, $D = 1$. Various entropy
distributions are shown, for which the bounds  for the  off-diagonal elements are equal,  $U=L$, 
and  are varied from $10^{-4}$ to $10^4$.  Note the  
logarithmic scale. 
\begin{figure}
\centering
   {\includegraphics[width=8cm]{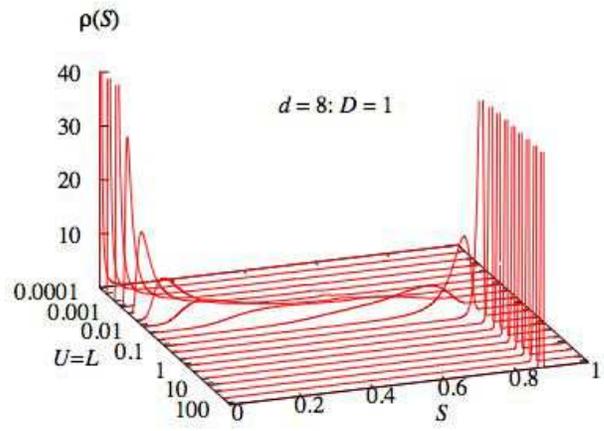}}
   \caption{(Color online)  Entropy distributions $\rho(S)$ in eight dimensions $(d = 8)$ for 
   random matrices
   with $D=1$ and upper bounds for the off-diagonal elements $U = L$ varying between $10^{-4}$    
   and unity.  The distributions are clipped for $\rho > 40$. The selected off-diagonal
   bounds $U=L$ are arranged on a logarithmic scale.}
 \label{fig3}
\end{figure}
The system  is clearly purifying if the off-digonal elements are small. However, it becomes mxing  for 
$U=L > 0.1$. This transition takes place for smaller values of $U=L$, if the dimension of the 
system is larger. This is demonstrated in Fig. \ref{fig3}  for $d = 8$.

\subsection{ Strong asymmetry: The dependence on $D$ and $L$ for $U=1$ }
\begin{figure}
\centering
   {\includegraphics[width=8cm]{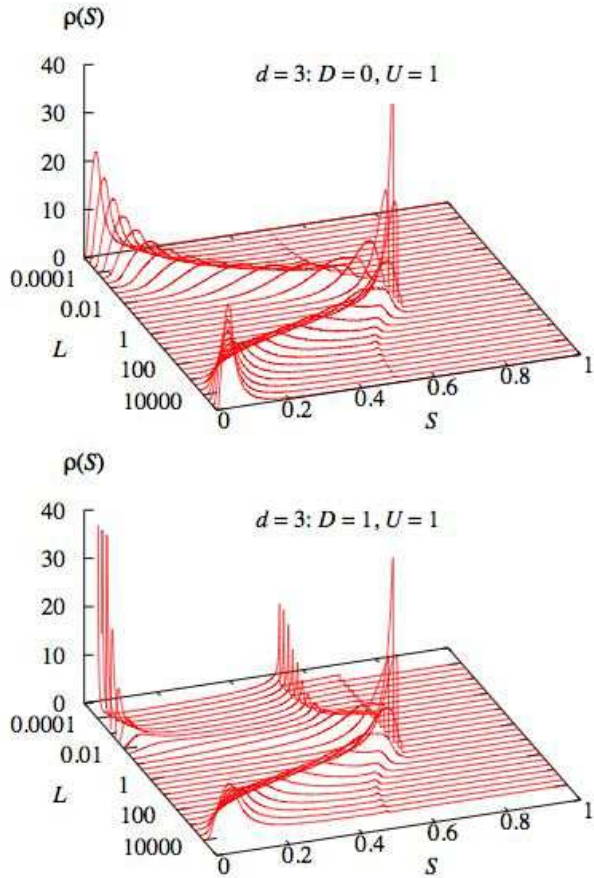}}
    \caption{(Color online)  Entropy distributions $\rho(S)$  in three dimensions $(d = 3)$ for
    various $L$ (logarithmic scale). The bound $U=1$,  whereas $D$ for the diagonal elements 
    vanishes at the top, and is unity at the bottom. 
   The distributions are clipped for $\rho > 40$.}
    \label{fig4}
\end{figure}

We stick with three dimensions and consider the case where
the bound for the upper off-diagonal elements is fixed, $U = 1$, whereas the 
respective bound for the lower off-diagonal elements is varied in steps over a huge range,
$10^{-5} \le L \le 10^{5}$. The top and bottom figures of Fig. \ref{fig4} differ by the choice of
$D$, the upper bound of the diagonal elements. In the top figure, $D$ is set to vanish, whereas in the
figure at  the bottom $D$ is set to unity. Although the off-diagonal
elements may be small, they are still strictly positive and the conditions of Section \ref{sec_equation} still apply.
One infers from the top of Fig. \ref{fig4} that for vanishing $D$ the system
- very gradually - becomes more purifying, if the ratio of upper bounds $U/L$
becomes large or small. In the limit $L = 0 $ the system is strictly purifying, although the 
convergence towards this state is slow as we have verified by direct integration of the 
equations of motion (\ref{eq_fisher}).  

The entropy distributions very often have peaks which belong to  mixed substructures  with a 
lower dimension (such as at $S = 0.5$ in  Fig. \ref{fig4}). This feature may be
observed  with many other examples given in this paper. 

\begin{figure}
\centering
   {\includegraphics[width=8cm]{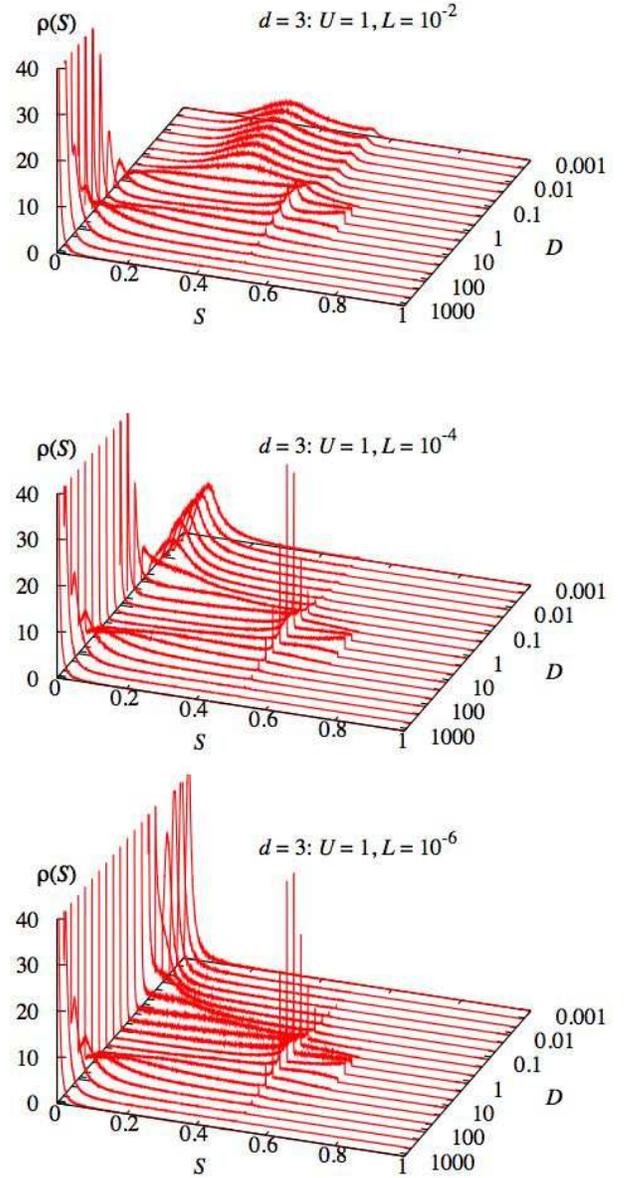}}
    \caption{(Color online)  Entropy distributions $\rho(S)$  in three dimensions $(d = 3)$ for
    various $D$ (logarithmic scale). The bound for the upper off-diagonal elements  $U=1$,
      whereas the respective bounds for the lower off-diagonal elements are $L = 10^{-2}$, 
      $10^{-4}$, and
     $10^{-6}$ for the top, middle and bottom figures, respectively.
   The distributions are clipped for $\rho > 40$.}
    \label{fig5}
\end{figure}

   To complement these results for three-dimensional systems, we show in Figure \ref{fig5} how the shape of the entropy distribution depends on the
upper bound of the diagonal elements $D$. Again, $U = 1$,
whereas the upper bound for the lower off-diagonal elements, $L$, is  $10^{-2}$,   $10^{-4}$, and
$10^{-6}$ for the figures from top to bottom, respectively.   
The surprising result is that for already rather small (but strictly positive) matrix elements below
the diagonal ( for $L = 10^{-2}$ at the top of Fig. \ref{fig5}), the system is mixing for small and vanishing 
(not shown) $D$, where the rather broad maximum of the  entropy distribution  only
very slowly moves towards the purifying case $S = 0$  for much smaller values of $L$.
(middle and bottom figure). 

\begin{figure}
\centering
   {\includegraphics[width=8cm]{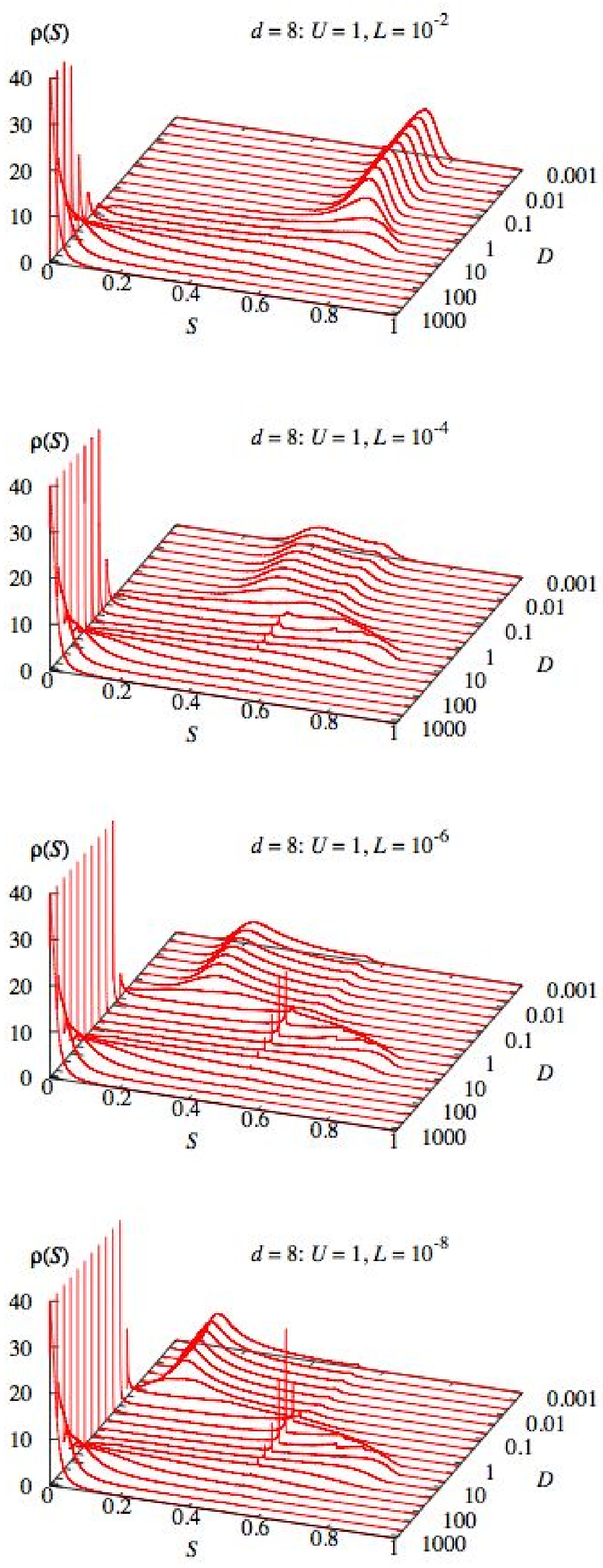}}
    \caption{(Color online)  Entropy distributions $\rho(S)$  in eight dimensions $(d = 8)$ for
    various $D$ (logarithmic scale). The bound for the upper off-diagonal elements  $U=1$,
      whereas the respective bounds for the lower off-diagonal elements are, from top to bottom, 
      $L = 10^{-2}$, $10^{-4}$, $10^{-6}$,  and $10^{-8}$.
   The distributions are clipped for $\rho > 40$.}
    \label{fig6}
\end{figure}

      This results also applies for higher dimensions. In Fig. \ref{fig6} we show 
the eight-dimensional analogue to Fig. \ref{fig5}, where again $\rho(S)$ is shown for 
various $D$, where $U$ is unity, and $L$ varies from $10^{-2}$ to $10^{-8}$
for the figures from top to bottom. Only for $U=0$ the system becomes strictly purifying for small $D$.

\section{The two-dimensional case}
\label{sec_theory}

\begin{figure}
\centering
   {\includegraphics[width=8cm]{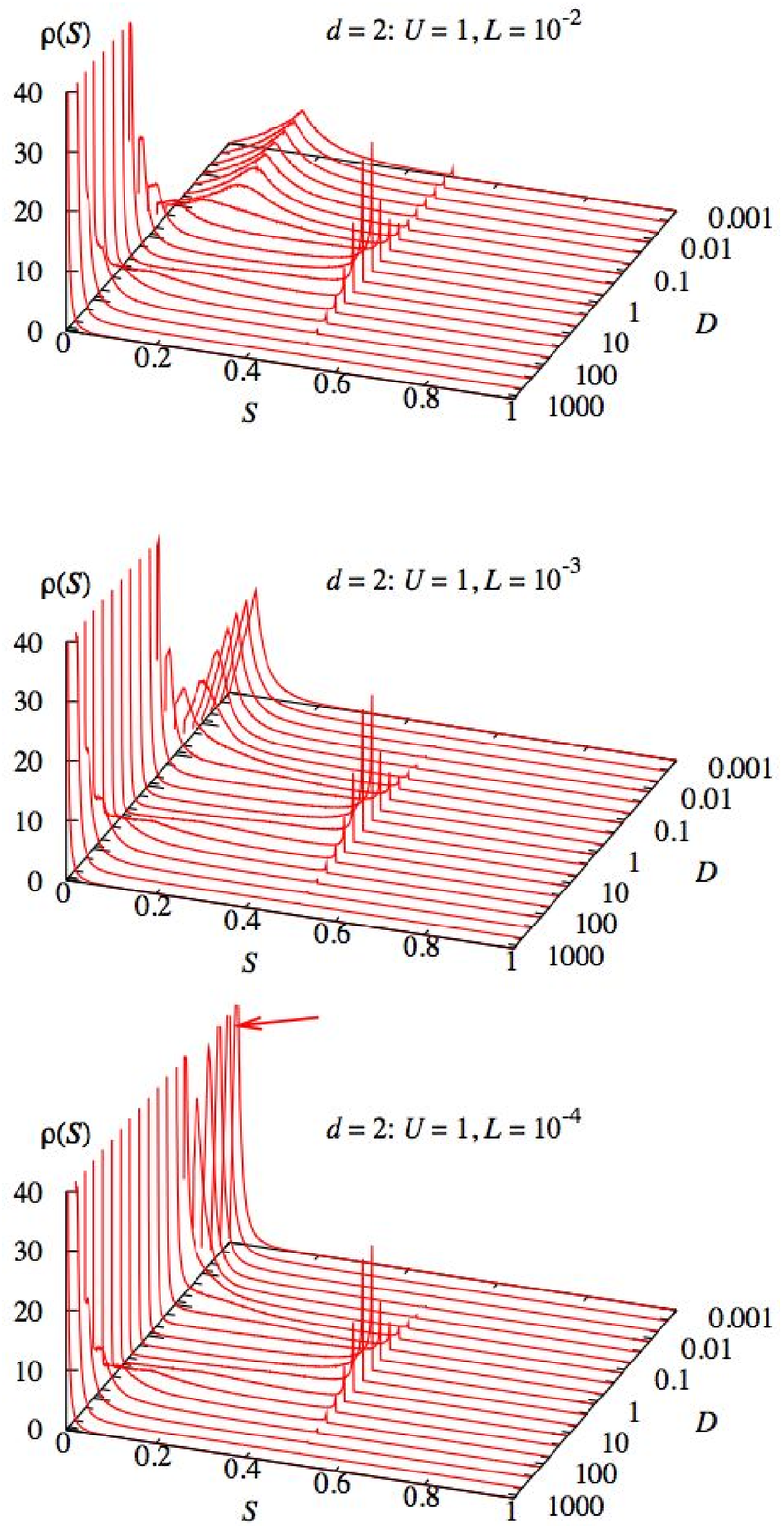}}
    \caption{(Color online)  Entropy distributions $\rho(S)$  in two dimensions $(d = 2)$ for
    various $D$ (logarithmic scale). The bound for the upper off-diagonal element  $U=1$,
      whereas the respective bounds for the lower off-diagonal element are $L = 10^{-2}$, 
      $10^{-3}$, and
     $10^{-4}$ for the top, middle and bottom figures, respectively.
   The distributions are clipped for $\rho > 40$.}
    \label{fig7}
\end{figure}

In an attempt to understand some of the details of this complicated behavior, we turn to the two-dimensional case, $d=2$, which may be solved analytically, at least in principle. The behavior
in two dimensions is qualitatively the same as for higher dimensions. This is demonstrated by the simulation results in 
Fig. \ref{fig7}, where entropy densities are shown for various $D$ varying over many orders of magnitude, and for bounds $L$, decreasing, from top to bottom, from $10^{-3}$ to $10^{-5}$,
respectively.
$S$ is bounded between 0 and 0.5 in this case.

      The $2 \times 2$ matrix of rate constants is written
\begin{equation} 
       \alpha = \begin{pmatrix} a & b \\ c & d \\    
\end{pmatrix} ,
\label{d2matrix} 
\end{equation}
where, in accordance with Eq. \ref{DUL}, the strictly positive elements are taken to be bounded by
\begin{equation}
0 < \{a,d\} < D \; ; \: 0 < b < U \: ; \; 0 < c < L .
\label{bounds}
\end{equation}
The maximum eigenvalue of $\alpha$ is 
$$
    \lambda = \frac{a+d}{2} + \sqrt{\left(\frac{a-d}{2}\right)^2 + b c}  .
$$    
According to Sec. \ref{sec_equation}, the corresponding (properly normalized) eigenvector $p$ with components  $(p_1, p_2)$  gives the asymptotic state of the system. From the eigenvalue
equation $\alpha p = \lambda p$ it immediately follows that the ratio
\begin{equation}
 r \equiv \frac{p_1}{p_2} = \frac{1}{c}\left[ \sqrt{ \beta^2  + b c} + \beta \right] ,
\label{constraint}
\end{equation} 
with $\beta = (a-d)/2$, only depends on the difference of the diagonal elements. 
The bounds for $\beta$ follow from those for $\{a,b\}$, Eq. \ref{bounds},
$ -D/2 < \beta < D/2 . $            
Since $a$ and $d$ are assumed to be uniformly distributed, $\beta$ has a triangular distribution,
\begin{equation}
    \pi(\beta) = \frac{2}{D}\left(1 - \frac{2}{D} |\beta| \right)\;;
     \int_{-D/2}^{D/2} \pi(\beta) d\beta = 1.
\label{pidistribution}
\end{equation}

The ratio $r = p_1/p_2$  is another indicator for mixing: A pure state corresponds to 
$r = 0$ and $r = \infty $, and the maximally mixed state has $r = 1$. $r$ is related to the
 quadratic entropy by 
$$
        S(r) = \frac{2 r}{(1+r)^2 } = S\left(\frac{1}{r}\right) .
$$
Once the normalized distribution for $r$, $\rho(r)$, is known, the corresponding entropy distribution
$\rho(S)$ follows from 
\begin{eqnarray}
\rho(S) = \frac{\rho(r(S))}{\left| d S(r) / dr \right| }  
              &=& \frac{\left(1 + \sqrt{1 - 2S} \right)^2}{2 S^2 \sqrt{1 - 2 S}}
                \rho(r_{+}(S)) \nonumber \\&+& \frac{\left(1 - \sqrt{1 - 2S} \right)^2}{2 S^2 \sqrt{1 - 2 S}}
                \rho(r_{-}(S)) .
\label{Jacobian}                
\end{eqnarray}       
Here, $r_+$ is used for the branch  $1 \le  r <  \infty$, and $r_-$ for the branch $0 < r < 1$:
$$
r_{\pm}(S) = \frac{1}{S} \left(1-S \pm \sqrt{1 - 2 S} \right).
$$
The distribution for $r$ is obtained from
\begin{eqnarray}
\rho(r) &=& \frac{1}{U L} \int_{0}^{U} d b \int_{0}^{L} d c \int_{-D/2}^{D/2} d \beta
     \nonumber \\
     & & \times    \pi(\beta)   \delta\left(r - \frac{\sqrt{\beta^2 + bc} + \beta }{c}\right) .
\label{rhor}
\end{eqnarray}        
Here, the $\delta$-function takes care of the constraining relation Eq. \ref{constraint} 
between $r$ and the  matrix elements $b, c$ and $\beta$, and  may be rewritten according to
\begin{eqnarray}
 \delta\left(r - \frac{\sqrt{\beta^2 + bc} + \beta }{c}\right) & = &
 \frac{1}{2r} \left(\frac{b}{r} + r c \right) \nonumber \\
                     &   & \times \delta \left( \beta - \frac{1}{2} \left(\frac{b}{r} - r c \right) \right).
 \nonumber 
 \end{eqnarray} 
Without loss of generality we consider $r \equiv r_+ > 1$
Since
$$
     \int_{-D/2}^{D/2} d \beta f(\beta) \delta( \beta - x ) = f(x)\Theta\left(\frac{D}{2} - |x| \right),
$$
where $\Theta$ is the unit step function, the distribution for $r$ becomes
\begin{eqnarray}
\rho(r) &=& \frac {1}{ 2 r U L} \int_0^U d b \int_0^L d c 
                \nonumber \\ &\times& \pi\left(\frac{1}{2} \left(r c - \frac{b}{r} \right) \right) 
                \left( r c +\frac{b}{r} \right)
                \Theta \left( D - \left| r c - \frac{b}{r}  \right|  \right) .
\nonumber
\end{eqnarray} 
With the transformation
\begin{equation}
        x = b/D r \;  ;  \; y =  c  r/D
\nonumber
\end{equation}
one obtains
\begin{eqnarray}
\rho(r) & = & (r x_0 y_0)^{-1} \int_0^{x_0} d x \int_0^{y_0} d y (1 - |y - x|)(x+y)  \nonumber \\
             &    &   \times \Theta(1 - |y-x|),
\label{finalrho}
\end{eqnarray}
where $x_0 \equiv U/D r$ and $y_0 \equiv L r/D$. The domain 
for the remaining integration in the $xy$ plane is the intersection of a rectangle, 
defined by $0< x < x_0; 0 < y < y_0$, and of a strip given by 
$-1 + x  <  y < 1+x$. For the general case this domain is slightly involved.
Here, we confine ourselves to a specific  situation, which is readily solved.  For example, we take
 $U = 1$, $L = 10^{-4}$, and $D=10^{-3}$. The simulation results  
 for this particular case  are marked by an arrow 
in the bottom panel of Fig. \ref{fig7}. Under these conditions,  $\rho(r)$ becomes a maximum
for $r =r_0 \equiv \sqrt{U/L} > 1$, for which $x_0 = y_0$ and the overlap of the rectangle with the
strip is maximized. 

\begin{figure}
\centering
   {\includegraphics[width=8cm]{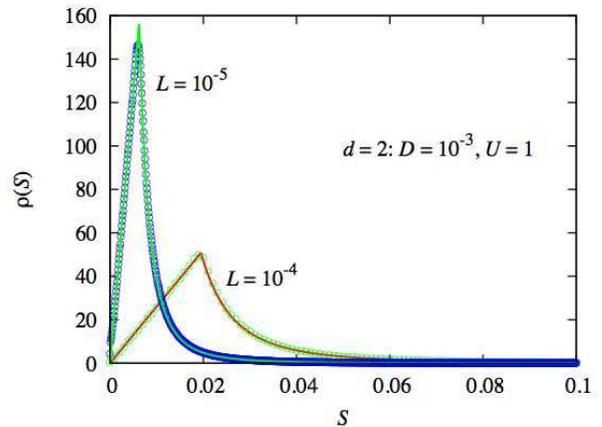}}\\
    \caption{(Color online)  Entropy distributions $\rho(S)$  in two dimensions $(d = 2)$ for
    $U=1$ and  $D = 10^{-3}$. As indicated by the labels, the two distributions are for
    $L = 10^{-4}$ (also marked with an arrow in the bottom figure of Fig. \ref{fig7}) and   
    $L = 10^{-5}$. The points are simulation results, and the smooth
    lines are the theoretical predictions as described in the main text.}
     \label{fig8}
\end{figure}

    For the regime of very large $r$,  $r = r_+ > r_0 > 1$, a direct
 evaluation of Eq. (\ref{finalrho}) gives $\rho(r) = (U/L) r^{-3}$. Since only $r_+$
 contributes, the entropy distribution is given by the first term of Eq. (\ref{Jacobian}). To lowest order in 
 $S$, one finds $ \rho(S) \approx (U / 4L) S $ for $S < S_0 \equiv 2 r_0/(1 + r_0)^2$,  which agrees well
 with the numerical results.    This is shown in Fig. \ref{fig8}.

   For the regime of slightly smaller $r$, $r_0 > r = r_+ > 1$, a similar evaluation of the integrals 
in   Eq. (\ref{finalrho})
yields $\rho(r) = (L/U) r $ and, hence, $ \rho(S) \approx  (4L/U)/S^3$ for $S > S_0$.  Again, this is in 
complete agreement with the numerical data (Fig. \ref{fig8}).       

   We conclude that all the details of the two-dimensional entropy distributions in Figure \ref{fig7} 
 are accessible to analytical computation. This is of some significance because of the qualitative similarity with the results for higher dimensions.
      
 \section{Fitness}
\label{sec_fitness}
 
 \subsection{The classical case}

If the evolution equations of Eq.  \ref{eq_fisher} are written as
$$
      \dot{p}_i = \sum_j \alpha_{ij} p_j - \phi p_i,
$$           
the global fitness of the state,
$$
\phi = \sum_j \sum_k \alpha_{jk} p_k
$$
evolves according to
$$
\dot{\phi} = \sum_j \sum_k \alpha_{jk} \left( \alpha_{kl} p_l - p_k \sum_l\sum_m 
  \alpha_{lm} p_m \right)
$$
It increases if $\alpha$ is  diagonal.

\subsection{The quantum setting}

Our results so far can readily be generalized to the quantum mechanical 
setting \cite{Alicki}, where the distribution $p_1, p_2,\cdots, p_d$  is replaced by a density
matrix $\rho$  which  obeys 
\begin{equation}
d\rho(t)/dt = h \rho(t) + \rho(t) h^{\dagger} - 
                 \rho(t) \mbox{Tr}(h \rho(t) + \rho(t) h^{\dagger} ). 
\label{quantumevol}
\end{equation}
$h$ is a sort of Hamilton operator, and $h^{\dagger}$ is its Hermitian adjoint. 
The solution is almost the same as 
in the classical case before,
$$
\rho(t) = \exp(ht) \rho(0) \exp(th^{\dagger}) [\makebox{Tr} \exp(ht) \rho(0) \exp(th^{\dagger})]^{-1},
$$
 just observe the non-commutativity of the operators. Entropy $S$ and fitness $F$ generalize to
$$
S = \makebox{Tr}  \rho(1- \rho); \;    F =  \makebox{Tr} \rho(h + h^{\dagger}) = dc/dt,
$$
 with 
 $$
 c(t)= \ln [\makebox{Tr} \exp(h t) \rho(0) \exp(t h^{\dagger})].
 $$
Thus, the increase in fitness means convexity of the function $c(t)$.
The increase of $F$ for diagonal $\alpha$  in the classical case becomes now
the following \\
\noindent
{\bf Proposition}: \\
$F$ is increasing if $h$ is Hermitian.\\
\noindent
{\bf Proof}:\\
The evolution  equation (\ref{quantumevol}) tells us generally
$$
dF/dt  = \makebox{Tr} \dot{\rho} (h + h^{\dagger}) = \makebox{Tr} [h \rho + \rho h^{\dagger} - 
                   \rho  \makebox{Tr} \rho (h + h^{\dagger})](h + h^{\dagger}) .
$$
For Hermitian $h$ this becomes four times $\makebox{Tr} (h^2 \rho) - (\makebox{Tr} \rho h)^2 $ .
The generalization of the Schwartz inequality to traces of operator products implies
$$
(\makebox{Tr} \rho h)^2  = (\makebox{Tr}  \rho^{1/2}  \rho^{1/2} h)^2 \le (\makebox{Tr}
        \rho ) (\makebox{Tr} \rho h^2),
$$
 which gives  $\dot{F}  \le 0$ for all $\rho$.

\noindent  
{\bf Remarks}: 

1) Without Hermiticity, $F$ need not increase. This appears already for a 
nilpotent $h$, $h^2 = 0$. Then $\exp(ht) = 1+ht$ and 
$$
\ln[\makebox{Tr}(1+ht) \rho(1+ th^{\dagger})] = \ln \makebox{Tr} [(1+th+ th^{\dagger} +
     t^2 h h^{\dagger}) \rho]
$$ 
is no longer convex in $t$. 

2) The Schwarz inequality becomes an equality only if the two vectors are 
proportional.  Therefore the proof tells us that the fitness keeps 
increasing unless $\rho$ projects onto an eigenvector of $h$. Of these, 
only the one with the highest eigenvalue has the biggest $F$ and, therefore, is an 
attractor.  Excluding 
degeneracy this means that then also the entropy goes to zero and the system becomes pure.
$S$ cannot be a Lyapunov function, because near a repellor it will be very small, so it has to increase
only to decrease again near the attractor.

3) The situation is just the converse for the Lindblad dynamics \cite{Narnhofer}, which provides a
consistent description of a dissipative quantum system \cite{Lindblad,LBW2001,Benatti}: 
\begin{equation}
\dot{\rho} = h \rho h^{\dagger} - \frac{1}{2} \left(h h^{\dagger} \rho + \rho h h^{\dagger}\right) .
\label{quantum_LB}
\end{equation}
This equation is linear in $\rho$ but quadratic in $h$. There, the fitness 
$F = \mbox{Tr} h \rho h^{\dagger}$ increases for a nilpotent $h$, 
$\dot{F} =  h^{\dagger} h \rho h h^{\dagger} \ge 0$, and is constant for $h = h^{\dagger}$.

\section{Concluding remarks}

The motivation for this study was to see, whether the nonlinear evolution
equations let order emerge from chaos purely by accident. The answer
depends where we let chance act.
Our results can be simply stated as follows: According to Eq. (\ref{DUL}) the matrix 
space is spanned by three parts, the diagonal matrices (D), and the upper (U)  and 
lower (L) triangular matrices. Matrices taken from solely one of these 
subalgebras give a purifying dynamics, but admixture of a 
small fraction of another part renders the dynamics mixing. 
Since zero is not a random number, these subalgebras are never pure,
but we keep the admixtures so small that they do not matter.
These features can be easily understood in the financial 
setting. There, the $p_i$ are the assets of party $i$, and the 
rules of the game are that per unit time party $k$  gives 
the ratio $\alpha_{ik}$ of its asset to party $i$. This money is 
not deducted from its account since the   $\alpha_{ik}$ are strictly positive,
but, instead, there is a flat 
tax which keeps the total amount of money fixed. The diagonal 
elements correspond to the case where each party makes its 
own investments and eventually the one with the highest 
interest rate becomes the richest. The triangular dynamics 
means that there is an order between the parties, and money is 
handed only from parties of lower order to the ones of 
higher order. Obviously, the one on top of this order pyramid 
eventually wins.

In a mixed dynamics these tendencies will compete, and since 
there are so many more possibilities for them to act in 
different directions, a random choice of mixtures will lead to 
a mixing dynamics.

Thus, within this equation only the simplest and most brutal 
strategies lead to the dominance of one party, but {\em fortuna} 
may easily even this out.

     We are grateful to Sumiyoshi Abe for a discussion on entropies and for pointing out
that the quadratic entropy was already used by E. Fermi as is mentioned in Ref. \cite{Jumarie}.
We also thank Reinhard B\"urger, Josef Hofbauer, Martin Nowak, 
Peter Schuster, Karl Sigmund and Jakob
Yngvason for stimulating discussions and useful comments.

\end{document}